\documentstyle[sprocl,epsf,calrsfs]{article}

\newcommand{\la}[1]{\label{#1}}
\newcommand{\nr}[1]{(\ref{#1})}
\newcommand{\nn}{\nonumber}
\newcommand{\lsi}{\raise0.3ex\hbox{$<$\kern-0.75em\raise-1.1ex\hbox{$\sim$}}}
\newcommand{\gsi}{\raise0.3ex\hbox{$>$\kern-0.75em\raise-1.1ex\hbox{$\sim$}}}

\begin{document}

\title{REAL TIME CORRELATION FUNCTIONS AT FINITE TEMPERATURE AND THEIR
  CLASSICAL LIMIT \normalsize{\footnote{ 
Talk given at the E\"otv\"os Conference in Science:\ Strong and
Electroweak Matter, Eger, Hungary, 21-25 May 1997
}}}

\author{D.~B\"ODEKER\footnote{bodeker@thphys.uni-heidelberg.de}}

\address{Institut f\"ur Theoretische Physik, Universit\"at Heidelberg, \\
Philosophenweg 16, 
D-69120~Heidelberg, Germany}


\begin{flushright}
HD-THEP-97-43\\
August, 1997
\end{flushright}
\vspace{0.8cm}

\maketitle 
\abstracts{
In order to investigate the reliability of the classical approximation
for non-perturbative real time correlation functions at finite
temperature we study the two-point correlator for the anharmonic
oscillator. For moderately large times the classical limit gives a
good approximation of the quantum result but after some time $t_*$ the
classical approximation breaks down even at high temperature.
}
  
\section{Introduction}

Real time processes at high temperatures play an important role in
the physics of the early universe and of heavy ion collisions.  Some
of the corresponding physical quantities, like the sphaleron rate, can
be related to real time correlation functions in finite temperature
field theory. A lot of work on this subject has been done in
perturbation theory. However, some physical quantities of interest are
non-perturbative and the only known method of calculating them on the
lattice is the classical approximation\cite{grigoriev}.


The usual argument why the classical approximation should be reliable
goes as follows: Typically, the non-perturbative effects are due to
field modes with small spatial momenta $|{\bf k}|\ll T/\hbar$. For
these momenta the Bose-Einstein distribution function is large,
$n(|{\bf k}|)\gg 1$. In this limit one expects physics to be
classical.

The question I would like to address in this talk is what one can say
about the reliability of the classical approximation in quantitative
terms. Is there an approximation scheme which tells us how large
quantum corrections to the classical approximation are?

There is such an approximation scheme for equal time correlation
functions called dimensional reduction\cite{laine}.  Equal time
correlation functions in ($d$+1) dimensions can be computed in an
effective classical theory which is a $d$-dimensional Euclidean field
theory.  Corrections to the effective theory can be estimated and they
are small at sufficiently high temperatures.

One possibility to investigate this problem for the case of unequal time
correlation functions is to expand the quantum correlation function
$C(t)$  in powers of
Planck's constant\cite{bodeker96},
\begin{eqnarray}
  \label{hbarexp}
  C(t)  = C_{\rm cl}(t)  + \hbar C_{1}(t) + \hbar^2  C_{2}(t)
   + {\cal O}(\hbar^3).
\end{eqnarray}
Each term in this expansion contains all orders in the coupling
constant.  The reliability of the classical approximation can then be
estimated by comparing the classical correlation $C_{\rm cl}(t)$
function with the first quantum correction.

\section{The Anharmonic Oscillator}

The anharmonic oscillator is the simplest interacting theory and it
can be viewed as a (0+1)-dimensional field theory. We
consider\cite{bodeker97} the finite temperature correlator
\begin{eqnarray}
  \label{cq}
 C(t)=
 \left\langle 
   \frac{1}{2}\Big[
   Q(t)Q(0)+Q(0)Q(t)
   \Big]
 \right\rangle ,
 \la{c}
\end{eqnarray}
where $\langle\cdots\rangle = Z^{-1}{\rm tr}[(\cdots)\exp(-\beta H)]$
denotes the thermal average and $Z={\rm tr}[\exp(-\beta H)]$.
Furthermore, $Q(t)=\exp(i H t/\hbar) Q(0)\exp(-i H t/\hbar)$ is the
position operator at time $t$. The Hamiltonian is $ H=P^2/2+U(Q)$ with
\begin{eqnarray}
  \label{u}
U(Q)=\pm \frac{1}{2}\omega^2Q^2+\frac{1}{4}g^2
Q^4 .
\end{eqnarray}
We refer to the two cases of a positive and of a negative quadratic
term as the symmetric and the broken case, respectively.

The classical approximation to Eq.~\nr{c} is given by
\begin{eqnarray}
  C_{\rm cl}(t)= Z_{\rm cl}^{-1} \int \frac{dp dq}{2\pi\hbar}
  e^{-\beta H(p,q)}  q q_{\rm c}(t) ,
  \la{ccl}
\end{eqnarray}
where $Z_{\rm cl} = \int dp dq/(2\pi\hbar) e^{-\beta H(p,q)}$
and $q_c(t)$ is the solution of the classical equations of motion
with the initial conditions $q_c(0)=q,$ $ \dot{q}_c(0)=p$. 

Expanding Eq.~\nr{c} in powers of $\hbar$ one finds that the lowest order term
is given by Eq.~\nr{ccl}. The term linear in $\hbar$ vanishes
and the first quantum correction to Eq.~\nr{ccl} is quadratic in
$\hbar$. Up to this order
\begin{eqnarray}
 C(t)\hspace*{-0.3cm} &=&  \hspace*{-0.3cm} Z^{-1} 
 \int \frac{dp dq}{2\pi\hbar}  
 e^{-\beta H(p,q)} 
  \biggl\{ \biggl[ 1  - \frac{\hbar^2\beta^2}{24}  U''(q) +
  \frac{\hbar^2\beta}{24} \left[
  \partial_q^2 
  +  U''(q) \partial_p^2 
 \right] \biggr] q q_{\rm c}(t)  \nn \\
&&\hspace*{2cm}  - \frac{\hbar^2}{24}
q    \int_0^t dt'  U'''\big(q_{\rm c}(t')\big) 
   \{q_{\rm c}(t'),q_{\rm c}(t)\}_3  \biggr\} + {\cal O}(\hbar^3), \la{cqc}
\end{eqnarray}
where $\{f,g\} = \partial_p f \partial_q g - \partial_p g\partial_q f$
denotes the Poisson bracket
and $\{f,g\}_0= g$, $\{f,g\}_{n+1} = \{f,\{f,g\}_n\}$.
Similarly, the expression for $Z$ to order $\hbar^2$ is
\begin{eqnarray}
Z= \int \frac{dp dq}{2\pi\hbar}  e^{-\beta H(p,q)} 
  \left[ 1  - \frac{\hbar^2\beta^2}{24} U''(q) \right]. \la{z}
\end{eqnarray}

The solutions of the classical equations of motion $q_{\rm c}(t)$ in
Eqs.\ \nr{ccl} and~\nr{cqc} can be written in terms of the Jacobi
elliptic functions. The phase space integration in
Eq.~\nr{ccl} cannot be done analytically. Computing the quantum
corrections in Eq.~\nr{cqc} is even more difficult.  Therefore we have
evaluated these expressions numerically and also in a large time
expansion. The latter can be applied only in the symmetric case. Since we
are interested in analytical results we will discuss  this case
here.  The numerical calculations and a detailed
discussion of the broken case can be found in Reference\cite{bodeker97}.
\begin{figure}[tb]
 
\vspace*{-0.5cm}
 
\hspace{-1.4cm}
\epsfysize=9cm
\epsffile{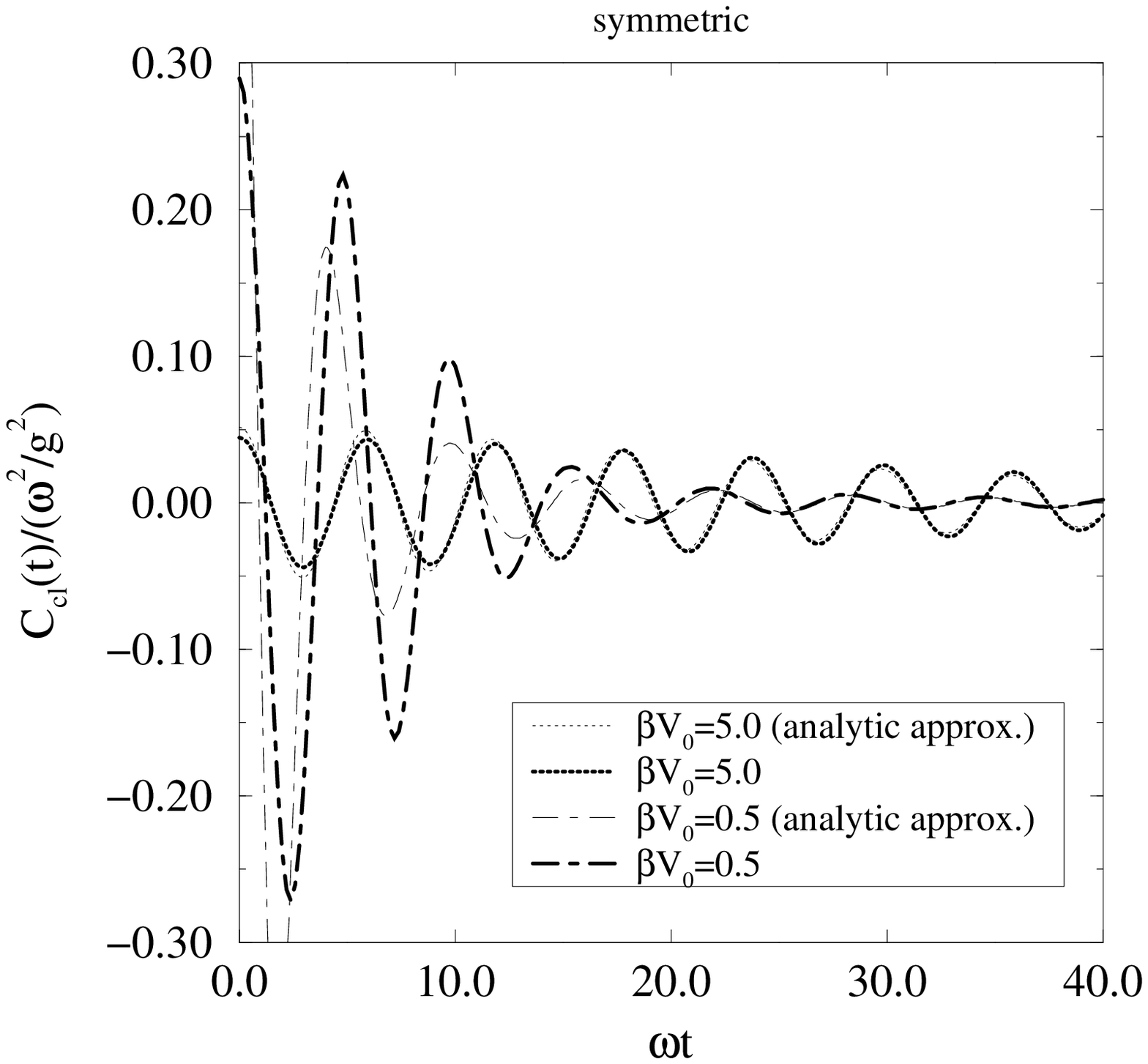}
\epsfysize=9cm
\epsffile{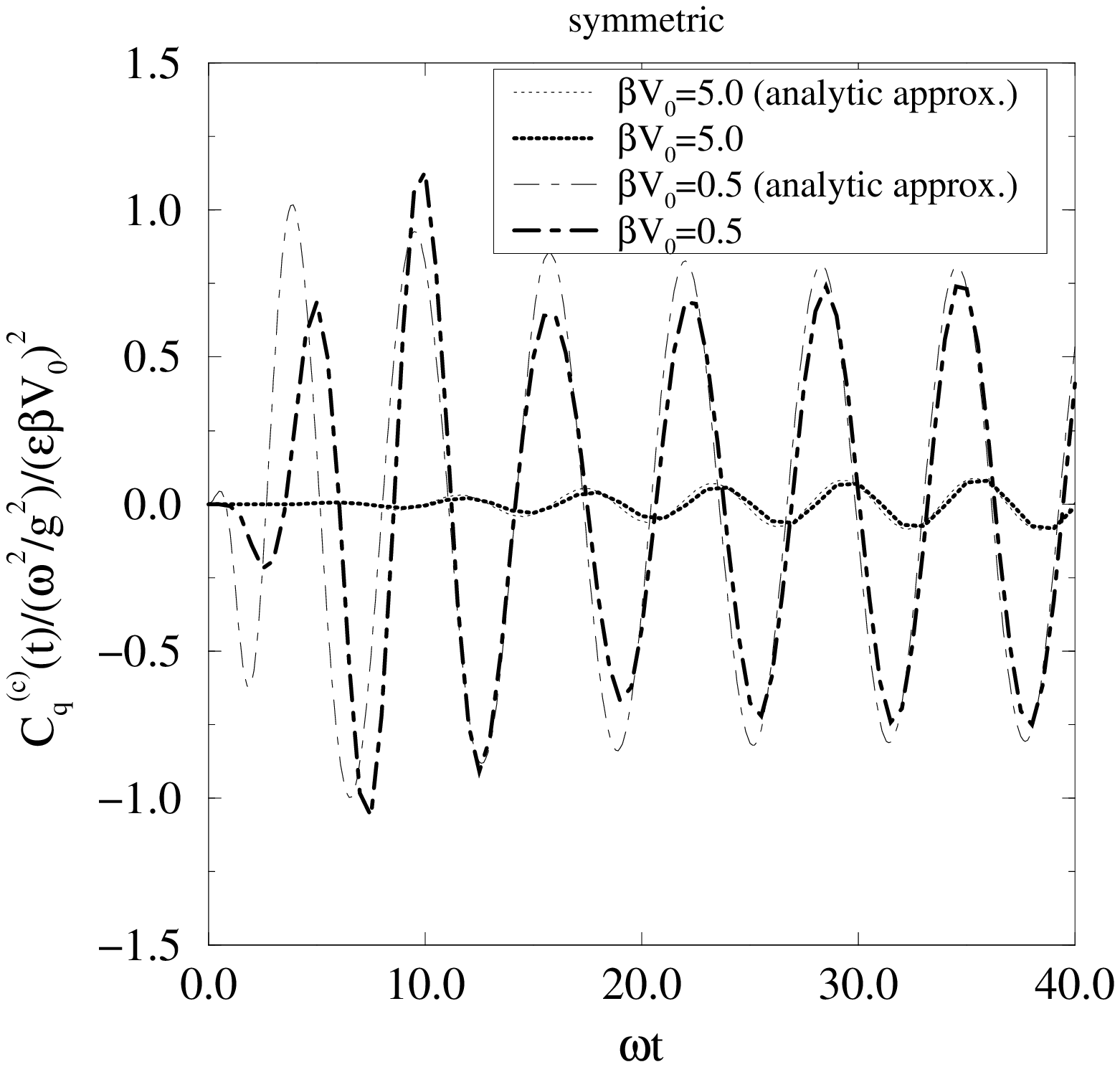}

\vspace*{-3cm}

\hspace{2.3cm} (a) \hspace{6cm} (b)
 
\caption[a]{
  (a) The classical correlator $C_{\rm cl}(t)$ and (b) the quantum
  correction $C_{\hbar^2}^{(c)}(t)$ in the symmetric case.  The thick
  lines represent the exact numerical results and thin lines represent
  the analytic approximation\cite{bodeker97}.  The dimensionless
  variables $V_0$ and $\epsilon$ are $ V_0=\omega^4/(4g^2)$,
  $\epsilon = g^2\hbar/\omega^3$.}

\label{figure}
\end{figure}

The results for the classical correlator $C_{\rm cl}(t)$ in the
symmetric case are  shown in Fig.~1(a). $C_{\rm cl}(t)$ is an
oscillating function.  Its amplitude attenuates since the frequency of
$q_c(t)$ is energy dependent and the different $q_c(t)$ in
Eq.~\nr{ccl} interfere destructively. For very large times
\begin{eqnarray}
  \label{asy}
  \omega t\gg 1, \quad \omega t \gg \beta \omega^4/g^2
\end{eqnarray}
the classical correlator can be approximated by
\begin{eqnarray}
  \label{cclasy}  
   C_{\rm cl}(t) \approx -  \frac{16}{9} \frac{1}{\hbar Z_{\rm cl}}
\frac{\omega^5}{g^4} \frac{\cos\omega t}{(\omega t)^2}.
\end{eqnarray}

Now we consider the order $\hbar^2$ corrections to $C_{\rm cl}(t)$
which we write as $\hbar^2 C_{2}(t) = C_{\hbar^2}^{(a)}(t) +
C_{\hbar^2}^{(b)}(t) + C_{\hbar^2}^{(c)}(t)$. The term
$C_{\hbar^2}^{(a)}(t)$ is a sum of the $\hbar^2$ correction to the
partition function when it combines with the classical result $C_{\rm
  cl}(t)$, and of the term proportional to $ U''(q)$ in the numerator
of Eq.~\nr{cqc}. $C_{\hbar^2}^{(b)}(t)$ contains the
terms proportional to $\partial^2_qq q_{\rm c}(t)$ and to
$\partial^2_pq q_{\rm c}(t)$.  Finally, $C_{\hbar^2}^{(c)}(t)$
is given by the term in Eq.~\nr{cqc} which contains the Poisson
brackets.

For large times $C_{\hbar^2}^{(a)}(t)$ and $C_{\hbar^2}^{(b)}(t)$ are
proportional to $\beta/(\omega t)^2$. At sufficiently high temperature
these quantum corrections are always small compared to the classical
approximation.  The results for $C_{\hbar^2}^{(c)}(t)$ are shown in
Fig.~1(b).  One sees that the amplitude does not decrease with time
and therefore $C_{\hbar^2}^{(c)}(t)$ will become larger than $C_{\rm
  cl}(t)$ at sufficiently large times. In the limit~\nr{asy}
\begin{eqnarray}
  \label{cqcasy}
   C_{\hbar^2}^{(c)} (t) \approx -  \frac{1}{12} 
\frac{1}{\hbar Z_{\rm cl}}
  \frac{\hbar^2}{\omega} \cos\omega t.
\end{eqnarray}
Comparing Eqs.\ (\ref{cclasy}) and  (\ref{cqcasy})
we see that $ C_{\hbar^2}^{(c)} (t)$ becomes as large as the classical
correlator for $t\sim t_*$ where
\begin{eqnarray}
 \omega  t_* = \frac{\omega^3}{\hbar g^2 },
 \la{t*}
\end{eqnarray}
and for $t>t_*$ the classical approximation breaks down.

So far we have considered the classical limit of Eq.~\nr{cq} without
modifying the parameters of the Hamiltonian.  One may wonder whether
one can ``improve'' the classical approximation by using an effective
theory in which the parameters are determined by dimensional
reduction. This would correspond to a resummation of certain quantum
corrections. 

To order $\hbar^2$ one can easily calculate the ``mass''
parameter of the effective theory:
\begin{eqnarray}
  \label{omeff}
  \omega_{\rm eff}^2=\omega^2+3g^2 T\sum_{n\neq 0}
  \frac{1}{(2\pi nT/\hbar)^2}=\omega^2+\frac{1}{4}g^2\hbar^2 \beta 
\end{eqnarray}
The change in the coupling constant is of order $\hbar^4$ and thus
does not contribute in the present $\hbar^2$-calculation.

Now consider Eq.~\nr{cqc}  were we have $U''=\omega^2+3
g^2q^2$. The term $\omega^2$ is canceled by the corresponding term in
$Z^{-1}$.  The term $3g^2 q^2$  in Eq.~\nr{z}
can be reproduced by calculating the classical partition function
$Z_{\rm cl}$ with $\omega_{\rm eff}$ instead of $\omega$.  Similarly,
the term $3g^2 q^2$  in Eq.~\nr{cqc} is accounted for by
using $\omega_{\rm eff}$. Thus $C_{\hbar^2}^{(a)}(t)$ can be taken
into account by changing the parameters of the classical theory.

There remain the terms $C_{\hbar^2}^{(b)}(t)$ and
$C_{\hbar^2}^{(c)}(t)$. We have seen that the latter is responsible
for the breakdown of the classical approximation.  The question is
whether it can be taken into account using $\omega_{\rm eff}$
in the classical equations of motion which determine $q_{\rm c}(t)$.
To answer this question we can simply replace $\omega\to\omega_{\rm eff}$
in Eq.~\nr{ccl} and expand in $\hbar$ to obtain
\begin{eqnarray}
  \label{cclres}
  C_{\rm cl}^{\rm eff}(t) \approx
\biggl[
1+b_1 \frac{g^2\hbar^2\beta}{\omega^2}\biggr] C_{\rm cl}(t)+
\frac29
\frac{1}{\hbar Z_{\rm cl}}
\frac{\hbar^2\beta\omega^3}{g^2} 
\frac{\sin\omega t}{\omega t} + {\cal O} (\hbar^3),
\end{eqnarray}
where $b_1$ is some number. Now we see that for large times $C_{\rm
  cl}^{\rm eff}(t)$  has a different functional behavior
than
$C_{\hbar^2}^{(c)}(t)$. 
Therefore the large quantum corrections due to
$C_{\hbar^2}^{(c)}(t)$ 
cannot be resummed in this way.

\section{Summary}

We have studied the classical finite temperature real time two-point
correlation function and its first quantum corrections for the
anharmonic oscillator.  The expansion around the classical limit is
made in powers of $\hbar$, so that each order contains all orders in
the coupling constant $g^2$. 

As long as $t \ll t_* = \omega^2/(\hbar g^2)$ the quantum corrections
are small compared to the classical correlation function. From this we
would expect that in this region the classical limit gives a good
approximation for the full quantum correlation function.  The
expansion parameter for the quantum corrections in this region is not
just the naive $(\beta\hbar\omega)$. Instead, also the factors
$\epsilon(\beta\hbar\omega)$ and~$\epsilon^2$, where $\epsilon = \hbar
g^2/\omega^3$, appear.
 
The semiclassical expansion breaks down at $t \sim t_*$ when the
quantum corrections become as large as the classical result. Moreover,
we found that these large corrections cannot be resummed by modifying
the parameters of the classical theory.

\section*{Acknowledgments}
I would like to thank   M.~Laine and O.~Philipsen
for enjoyable collaboration and 
for useful discussions.

\end{document}